# $45^0$ rotated epitaxial nucleation of diamond on silicon using chemical vapor deposition

Qijin Chen

*The State Key Laboratory of Surface Physics, Institute of Physics, Chinese Academy of Sciences, Beijing 100080, China*

*Department of Physics, The University of Chicago, 5720 S. Ellis Ave., Chicago, IL 60637*

$45^O$ rotated epitaxial diamond nucleation on both (001) and {111} planes of a silicon substrate was observed using chemical vapor deposition with nucleation enhancement by electron emission. The epitaxial relationship between diamond and Si is D(001)//Si(001) and D<100>//Si<110> in both cases. While the case of diamond nucleation on Si(001) may be explained in Verwoerd's model, the other case has no theoretical model as yet. The as-grown diamond was characterized by scanning electron microscopy and Raman spectroscopy. It is speculated that a suitable combination of low temperature and low $CH_4$ concentration leading to low nucleation rate favor the 45° rotated epitaxy over those in parallel registry. This observation sheds light on the mechanism of epitaxy of diamond on Si substrates, and has great significance in parameter control in achieving single-crystal epitaxial diamond films. © *1997 Qijin Chen @ The University of Chicago.*
[S0003-6789(97)12345-0]

Great progress has been made in diamond synthesis at low pressure in the past several years, especially in the subject of diamond epitaxy on single crystal silicon substrate using chemical vapor deposition (CVD). Early in 1990, Jeng *et al.*[1] reported local epitaxy of diamond on Si via microwave plasma CVD (MPCVD). Shortly after that, epitaxially oriented diamond films were reported on β-SiC and Si substrate via a β-SiC intermediate layer in MPCVD[2,3,4]. They have also been achieved using hot filament CVD (HFCVD) recently[5]. Epitaxy of diamond directly on Si without an intermediate layer has been confirmed through high resolution transmission electron microscopy (HRTEM)[6,7]. However, due to the large lattice mismatch between diamond and Si (lattice constant: Si 5.4301Å, diamond 3.5667Å), no single-crystal diamond film has been achieved as yet; all the oriented films reported are mosaic at best, composed of oriented micro-crystals with obvious mismatch and grain boundaries. Meanwhile, different misorientation angles have been observed between the epitaxial diamond and the Si substrate, resulting from the large mismatch[7].

Theoretical efforts have also been made on various models of the diamond/Si epitaxy. The large lattice mismatch gives rise to the viability of epitaxial models other than the simple parallel 1:1 registry, which is the most commonly observed for small mismatch systems. The ratio of the lattice constant (Si/diamond) is nearly 3:2, reminding a 3:2 registry, as proposed by Verwoerd[8] and reported by Jiang and Jia[9]. So far, all the observations and models share the same characteristics, i.e., the diamond lattice is approximately in parallel with that of Si, the substrate. Models of diamond epitaxy on Si(001) rotated by $45^0$ about the common [001] axis were also put forward as early as in 1991[10,11]. However, experimental evidence in favor of this model has not been reported as of today. As a result of lack of experimental support and stimulation, theoretical study has not progressed rapidly recently. Therefore, it is necessary to check this prediction to make clear the mechanism of diamond/Si epitaxy, as it is also of great help in achieving single-crystal diamond epitaxy in its own right.

The purpose of this article is to report the observation of $45^0$ rotated epitaxial diamond nucleation with (001) texture on both (001) and {111} planes of a chemically etched Si(001) wafer via HFCVD, and to investigate the mechanism of epitaxial diamond nucleation on Si. The nucleation was achieved using the electron-emission-enhancement technique[12,13]. This result confirms Verwoerd's model, bringing into attention the critical issue on how to control deposition parameters to achieve single-crystal diamond films. It also presents the need of a theoretical model to explain the (001) textured diamond nucleation on Si{111} interfacial plane.

Our experiment was carried out in an HFCVD apparatus as reported in ref. 13 under the configuration of Fig. 1 therein. To repeat briefly, a φ140 mm and 500 mm long fused silica tube was used as deposition chamber. A Mo plate was placed on a copper platform to support samples. Coils of φ0.2mm tungsten wires were used as filaments. A diamond film was pre-deposited on the Mo substrate holder, which was negatively biased, to emit electrons during the nucleation stage. The source gas was $CH_4$ diluted in hydrogen. The substrate was a chemically etched p-type (001) Si wafer. Before loaded into the chamber, the Si substrate was chemically cleaned with acetone in an ultrasonic bath for 10 min, followed by 1 min rinse in 30 vol. % HF solution. Experimental parameters are listed in Table I. The $CH_4$ concentration, the substrate temperature and the filament temperature were all relatively low as compared with normal conditions.

The purpose to choose the chemically etched Si wafer as the substrate was to show clearly the relationship between the lattice orientations of the diamond overgrowth and the silicon substrate. This can be seen from Fig. 1, the SEM image of the substrate before deposition. The (001) facets of

**TABLE I. Experimental conditions.**

| Parameters | Nucleation | growth |
|---|---|---|
| Flow rate (sccm) | 100 | 100 |
| $CH_4$ concentration (vol. %) | 0.6 | 0.6 |
| Pressure (torr) | 20 | 20 |
| Filament temperature (°C) | 1920 | 1920 |
| Substrate temperature (°C) | 700 | 670 |
| Bias voltage (V) | -280 | 0 |
| Emission current (mA) | 150 | 0 |
| Time (hours) | 1 | 14.5 |

Si are clearly shown as squares in the figure. Between neighboring squares are {111} facets, which indicate that the edges of the squares are in <110> directions, instead of <100> directions. Thus, the orientation relationship between epitaxial diamond nuclei and the substrate will be clearly shown, without using the pole figure of X-ray analysis.

Figure 2a and 2b are the SEM image of the sample after deposition. The left half of Fig. 2a is the image of the diamond film on the Si(001) square in the central region of the right half with 5 times higher magnification. Fig. 2b is the image of the film on another Si(001) square of the sample with even higher magnification, to show the shape and orientation of the diamond micro-crystals more clearly. The low $CH_4$ concentration and the low temperature led to very low nucleation and growth rate. As a result, the sample is not completely covered with diamond nuclei, and therefore, the Si(001) squares resulting from the chemical etch and the Si <110> directions are still clearly visible in the right half of Fig. 2a. The shape of the diamond particles is not very sharp. Nevertheless, their pyramid shape with (001) top and {111} side faces are clear enough to show their orientations. The edges of the top square of the diamond pyramids are in D<110> direction, where D denotes diamond. Most diamond crystals in Fig. 2b and the left half of 2a share the same spatial orientation, indicating a fixed epitaxial, rather than random, orientation relationship with the substrate. There is an obvious rotation of approximately 45° between diamond and Si about their common [001] axis. Therefore, the epitaxial relationship is D(001)//Si(001), while there exists a rotation angle between D<110> and Si<110>, namely, D<100>//Si<110>. The exact angle is difficult to obtain as a result of the low crystallinity, small size of the crystals and possible fluctuations of this angle. Furthermore, the outskirts of Si(001) squares are Si{111} surfaces. Therefore, the diamonds in the outer region of Fig.

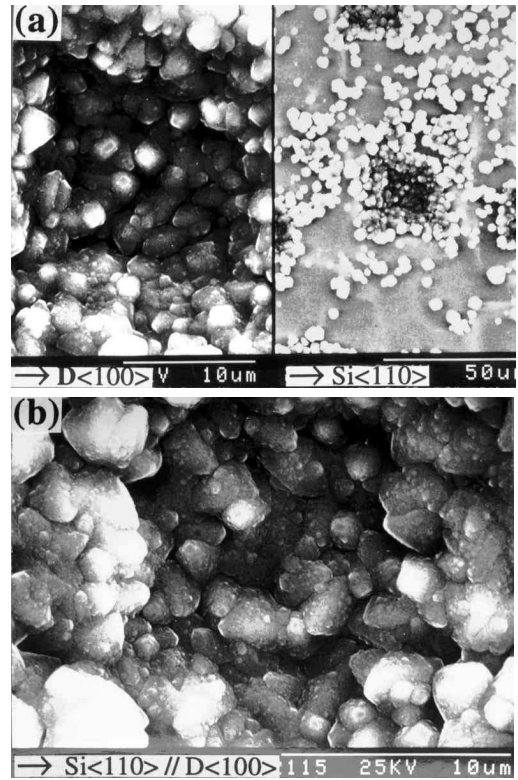

Fig. 2 SEM images of the sample after deposition. The left half of (a) is the central Si(001) square in the right half with a higher magnification. Fig. 2b shows epitaxial nuclei on another Si(001) square with even higher magnification. D<110> is rotated about the common [001] axis by 45° relative to Si<110>. In the outer region of (b), diamond nucleated on Si{111} planes. The 'um' under the scale bars denotes 'µm'.

2b and the left half of 2a actually nucleated on Si{111} planes. Still, they have the same epitaxial orientation with those that nucleated on the Si(001) square. Theoretical models for this kind of epitaxial diamond/Si interfacial structure are yet to come.

The Raman spectrum of the sample is shown in Fig. 3. The characteristic peak of diamond at 1332 cm$^{-1}$ is clearly shown, although the diamond crystallinity is not very good.

The ratio of the lattice constants between Si and diamond is 1.52, suggesting naively a parallel epitaxial interface structure of diamond on Si(001) in 3:2 registry, as put forward by Verwoerd[8]. He also proposed R45° structural models of diamond epitaxy on Si(001)[11], in which a 45° rotation about the common [001] axis is energetically favored. His calculations show that this rotation lowers the lattice mismatch between diamond and Si from 52% in the parallel registry to 7.3%, and reduces the interface energy to 2.07 eV per primitive Si surface unit cell. A revised model[8,14] with the introduction of a Si atom to the second diamond ad-layer per surface unit cell further reduces the interface energy to as low as 0.55 eV per surface unit cell. This makes the R45° configuration the most energetically favorable in comparison with various models in parallel 3:2 registry. Although other effects e.g. misfit dislocation, mismatch angle between [001] axes are not taken into account in the models above, it still indicates the large possibility of a R45° configuration. Our result serves as evidence of this R45° model.

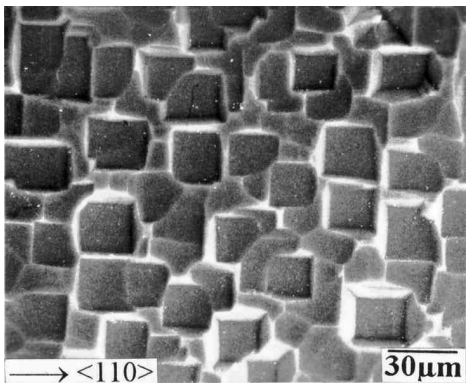

Fig. 1 SEM image of the chemical etched Si(001) sample before deposition. The squares are Si(001) facets with their edges in <110> directions. Between neighboring (001) squares are Si{111} facets.

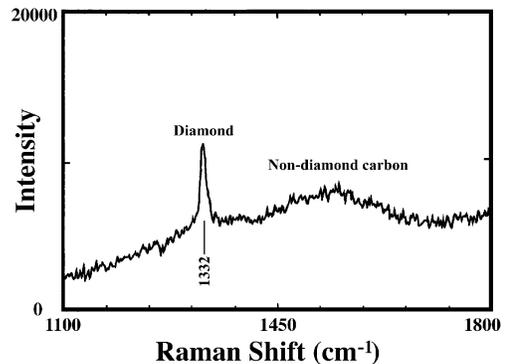

Fig. 3 Raman spectrum of the as-grown sample.

Although the revised R45° model has much lower interface energy[8,14], it does not necessarily mean that this is the case for our sample. As in Verwoerd's papers[8,11], more than one configurations of the interfacial structure give rise to the 45° rotation. To decide which situation applies to our experiment, HRTEM is needed to reveal the atomic interfacial structure. Unfortunately, this is limited by the resolution of HRTEM and the difficulty to get a good, observable interface sample for HRTEM. More efforts and technical progress have to be made to really make clear the interface structure. It is likely that there exists a small mismatch rotation angle between D[001] and Si[001] axes, just as HRTEM has revealed in other cases in parallel registry[6,7]. From the viewpoint of interface energy, the revised R45° model is the most probable case for our experiment.

On the other hand, the case in which diamond nucleated on Si{111} planes still awaits theoretical explanations. It is different from all other situations of epitaxy of diamond on Si reported up to date. In those cases, epitaxy of (111) diamond takes place on Si(111) plane, while (001) diamond nucleate on Si(001) plane. Our new result tells that more imaginative effort should be taken to reveal the mystery of the mechanism of diamond epitaxy on Si. Theoretical study needs to be done to investigate whether or not the R45° configuration but with a (111) interfacial plane is energetically favorable. It is possible that a {111} interface may relax the mismatch between the overgrowth and the substrate.

So far, various modes of epitaxy of diamond on Si have been reported. Yang et al[6] reported a 7.3° mismatch angle between D[001] and Si[001] axes in the case of diamond nucleation on Si(001) with 40 diamond {111} layers matching 25 Si{111} layers. The author and coworkers[7] reported a 9° angle in a similar situation, where 60 D{111} layers match 36 Si{111} layers. These ratios are close to 1.52, in an approximate parallel 3:2 registry. Jiang and Jia[9] observed epitaxy of diamond on Si(001) in 3:2 registry with a very small mismatch angle. Now we have an epitaxy of a 45° rotation. As far as Si(001) substrate is concerned, we have observed various different cases of direct epitaxy of diamond, not including the cases with β-SiC intermediate epitaxial layer. As already noted in earlier work[7], there exists competition among these different modes and with other non-epitaxial modes. Precise control such that only one of the epitaxial modes takes place is critical to synthesis of single-crystal diamond films on Si substrates.

While the sample pretreatment and the electron emission during the nucleation stage are essential to ensure epitaxial nucleation[5,12], what are the experimental conditions responsible for the R45° epitaxy? To answer this question, we note that, as part of the conclusions of ref. 14, the 45° rotated structure is particularly favorable for very thin diamond layers, while the parallel growth mode becomes most favorable for thick layers. There exists a stretch of almost 8% to match the substrate lateral lattice in the R45°, while 3:2 model only require 1.7%. In result, the adhesive binding energy (3.01 eV per surface unit cell) for the R45° model is lower than the 3:2 single bridge mode (3.63 eV per surface unit cell), though the latter has a much higher interface energy (3.09 eV per surface unit cell). The higher the binding energy, the easier to get a stable nucleus. Therefore, to get stable diamond nuclei or epitaxial thin layers in R45° mode, it requires a low nucleation rate to allow enough long time to convert thin layers in parallel modes to layers in the revised R45° mode, which is the most favorable before the layers grow thick. This requires a suitable combination of low substrate and filament temperature, as well as low $CH_4$ concentration to reduce the nucleation rate. In contrast, at normal conditions (filament temperature $\geq$ 2000°C, substrate temperature $\geq$ 800°C, $CH_4$ concentration $\geq$ 1.0 vol. %), the nucleation rate is much higher, thicker layers form rapidly within a period shorter than the time scale for parallel layers to convert to R45° layers, leading to epitaxy in parallel modes. Table I shows that our filament temperature, substrate temperature and $CH_4$ concentration were very low, so that the R45° mode won the competition against the parallel ones.

However, the reality may be much subtler. R45° nucleation did not dominate the entire sample, as can be seen from the right half of Fig. 2a. The Si(001) squares in the center forms an actual basin for the ad-species on the Si surface during the nucleation. It might have been a trap for reactive hydrocarbon species and changed their local concentration, etc., making it difficult to accurately describe the experimental conditions and to repeat the experiment successfully. This also renders the above argumentation speculative though it does sound plausible.

Effects of misfit dislocations and small mismatch angles were not taken into account in Verwoerd's model. They may relax the stretch and the mismatch between diamond and Si, and further lower the interface energy in both R45° and parallel modes. In any case, we still expect that the main results remain valid even after these considerations, else new models have to be constructed to accommodate our experimental result.

The possibility of an intermediate epilayer might also present. More theoretical and experimental work has to be done.

In summary, epitaxial nucleation of diamond of both Si(001) and Si{111} planes with the same epitaxial orientation relationship: D(001)//Si(001) and D<100>//Si<110>. There exists a 45° rotation of the diamond (001) plane about its [001] axis. While the former case may be explained with Verwoerd's revised R45° model, the latter is still awaiting theoretical investigation. It is speculated that a suitable combination of low substrate and filament temperature and low $CH_4$ concentration lead to low nucleation rate and thus favor the R45° epitaxial mode while normal conditions give rise to various parallel modes. This helps parameter optimi-zation in achieving mono-crystalline epitaxial diamond films on silicon substrates.

The author thanks Z.D. Lin for support for this work, Y.J.Yan, Q.L.Wu, and X. Kuang for operating the SEM. This work is financially supported by Chinese Natural Science Foundation, 863 Program and Beijing Zhongguancun Associated Center of Analysis and Measurement.